\documentclass[10pt,aps,prc, reprint,superscriptaddress,twocolumn,longbibliography,a4paper]{revtex4-1}
\usepackage{ulem}
\usepackage{graphicx}
\usepackage{soul}  
\usepackage{appendix}
\usepackage[dvipsnames,svgnames]{xcolor}
\definecolor{purple}{rgb}{0.3,0,0.9} 
\definecolor{darkteal}{HTML}{045D5D}
\usepackage[colorlinks=true,linkcolor= blue, citecolor = magenta, urlcolor = cyan ]{hyperref}
\hypersetup{breaklinks=true}
\usepackage{epstopdf}
\usepackage{amsmath,amssymb,amsthm,amsfonts} 
\usepackage{comment}
\usepackage{orcidlink}
\usepackage{mathrsfs}
\usepackage{xurl}
\newcommand{\be}{\begin{equation}}
\newcommand{\ee}{\end{equation}}
\newcommand{\ba}{\begin{eqnarray}}
\newcommand{\ea}{\end{eqnarray}}

\begin{document}
\title{Bottomonium suppression with a machine-learning-informed Debye mass}
\author{Ajaharul Islam\orcidlink{0000-0001-8174-907X}}
\email{aislam2@kent.edu}
\affiliation{Institute of Particle Physics and Key Laboratory of Quark and Lepton Physics (MOE), Central China Normal University, Wuhan, 430079, China}
\author{Shibo Chen\orcidlink{0009-0008-7915-5422}}
\email{ccc.physics@mails.ccnu.edu.cn}
\affiliation{Institute of Particle Physics and Key Laboratory of Quark and Lepton Physics (MOE), Central China Normal University, Wuhan, 430079, China}
\author{Fu-Peng Li\orcidlink{0000-0002-8763-385X}}
\email{fpli@fudan.edu.cn}
\affiliation{Key Laboratory of Nuclear Physics and Ion-beam Application (MOE) \&
Institute of Modern Physics, Fudan University, Shanghai 200433, China}
\affiliation{Shanghai Research Center for Theoretical Nuclear Physics,
NSFC and Fudan University, Shanghai 200438, China}
\author{Long-Gang Pang\orcidlink{0000-0002-1279-7008}}
\email{lgpang@ccnu.edu.cn}
\affiliation{Institute of Particle Physics and Key Laboratory of Quark and Lepton Physics (MOE), Central China Normal University, Wuhan, 430079, China}
\affiliation{Artificial Intelligence and Computational Physics Research Center, Central China Normal University, Wuhan 430079, China}
\author{Guang-You Qin\orcidlink{0000-0003-1173-1143}}
\email{guangyou.qin@ccnu.edu.cn}
\affiliation{Institute of Particle Physics and Key Laboratory of Quark and Lepton Physics (MOE), Central China Normal University, Wuhan, 430079, China}
\affiliation{Artificial Intelligence and Computational Physics Research Center, Central China Normal University, Wuhan 430079, China}
\begin{abstract}
Motivated by recent progress in data-driven approaches, we introduce a machine-learning (ML)-informed Debye mass, extracted from lattice-informed inputs, exclusively in the complex-valued heavy-quark Kent State University (KSU) potential. The resulting complex potential is used to solve the real-time Schr\"odinger equation within the quantum trajectories (QTraj) framework for the evolution of bottomonium in the quark-gluon plasma. We then compute the nuclear modification factors and double ratios for bottomonium $\Upsilon(1S)$, $\Upsilon(2S)$, and $\Upsilon(3S)$ states in Pb-Pb collisions at $\sqrt{s_{NN}} = 5.02$ TeV. We compare our ML-induced results with those from the original KSU model and with experimental measurements from ALICE, ATLAS, and CMS collaborations. We find that the machine-learned Debye mass leads to improved agreement with data, particularly for excited states, highlighting the utility of machine learning in modeling in-medium QCD effects.
\end{abstract}
\maketitle
\section{Introduction}
Ultra-relativistic heavy-ion collisions at the Relativistic Heavy Ion Collider (RHIC) and the Large Hadron Collider (LHC) have established the formation of a new state of strongly interacting matter known as the quark--gluon plasma (QGP), in which quarks and gluons are deconfined over nuclear volumes\cite{STAR:2005gfr, ALICE:2010suc}. Experimental measurements indicate that this medium behaves as a nearly perfect fluid with strong collective flow and a remarkably small shear viscosity to entropy density ratio. Understanding the microscopic properties of the QGP and the mechanisms governing its evolution remains one of the central goals of modern high-energy nuclear physics.

Heavy quarkonium states, bound states of heavy quark--antiquark pairs such as bottomonium ($\Upsilon(1S)$, $\Upsilon(2S)$, $\Upsilon(3S)$), provide particularly valuable probes of the QGP. Due to their large masses, heavy quark pairs are produced predominantly in the early hard scatterings of the collision and subsequently traverse the evolving medium. Their survival probability therefore carries information about the in-medium modification of the heavy-quark interaction and the thermodynamic properties of the QGP~\cite{Matsui:1986dk, STAR:2005gfr, Prino:2016cni}. The seminal work of Matsui and Satz \cite{Matsui:1986dk} proposed that color screening in a deconfined medium would suppress quarkonium binding, making quarkonium suppression a direct signal of deconfinement. Measurements at RHIC and the LHC have since revealed a characteristic sequential suppression pattern in which states with smaller binding energies melt at lower temperatures~\cite{PHENIX:2006gsi, ALICE:2014wnc, CMS:2018zza, CMS:2019uhm}. In particular, bottomonium suppression provides a cleaner probe charmonium because regeneration effects are expected to be small at LHC energies~\cite{Prino:2016cni, Du:2015wha, Andronic:2015wma}.

A theoretical description of quarkonium dynamics in the QGP requires accounting for both screening of the heavy-quark potential and in-medium dissociation processes. In this context, the heavy-quark potential becomes complex-valued,
\[
V(r,T)=V_R(r,T)+iV_I(r,T),
\]
where the real part governs the binding energy while the imaginary part encodes thermal broadening and decoherence arising from medium interactions. Effective field theory approaches~\cite{Brambilla:2004jw, Brambilla:2008cx, Rothkopf:2019ipj} and lattice QCD studies~\cite{Burnier:2014ssa, Datta:2003ww, Aarts:2011sm} have provided important insight into the structure of the in-medium heavy-quark potential. The imaginary component can be understood as originating from Landau damping and singlet--octet transitions induced by interactions with the surrounding medium~\cite{Brambilla:2011sg, Burnier:2014ssa}. These effects lead to a finite thermal decay width for quarkonium states.

Beyond static potential models, an important theoretical challenge is to describe the real-time quantum evolution of quarkonia in the QGP~\cite{Islam:2020gdv, Islam:2020bnp}. In recent years, open quantum system approaches have been developed to address this problem~\cite{Brambilla:2017zei, Akamatsu:2014qsa, Yao:2021lus, Hitschfeld:2024gtt, Yang:2024ejk, Brambilla:2022ynh, Brambilla:2023hkw, Strickland:2024oat, Brambilla:2024tqg, Brambilla:2025sis, Thapa:2025jua}. In particular, stochastic methods based on the Lindblad equation and the quantum trajectories (QTraj) framework~\cite{Omar:2021kra}, provide a practical way to model decoherence and dissociation dynamics of heavy quarkonium states propagating through a time-dependent medium. These approaches allow for a dynamical description of quarkonium suppression that naturally incorporates both screening and thermal decay effects.

Despite these theoretical advances, a key source of uncertainty in phenomenological descriptions of quarkonium suppression lies in the determination of medium parameters, particularly the Debye screening mass and the temperature-dependent strong coupling entering the heavy-quark potential. Traditionally, these quantities are obtained using perturbative approximations or phenomenological parameterizations. However, the strongly coupled nature of the QGP suggests that more data-driven approaches may provide improved constraints on these quantities.

In recent years, machine learning (ML) techniques have emerged as powerful tools for extracting physical information from complex datasets and solving inverse problems in physics through data-driven and physics-driven neural networks~\cite{RAISSI2019686, 712178, KHOO_LU_YING_2021, ZHOU2024104084, Pang:2016vdc, Aarts:2025gyp, Pang:2024kid}. In the context of QCD matter, deep neural networks (DNNs) have been successfully applied to reconstruct the QCD equation of state and infer quasi-particle properties from lattice QCD thermodynamics. By embedding physical constraints into the learning process, such models can provide data-driven estimates of medium properties that remain consistent with underlying theoretical principles.

Motivated by these developments, in this work we incorporate machine-learning--derived medium properties into the modeling of quarkonium suppression. In particular, we employ a deep-learning quasi-parton gas framework to obtain temperature-dependent effective coupling and Debye screening masses consistent with lattice QCD thermodynamics. The resulting machine-learning--informed Debye mass is introduced into the imaginary part of the complex Karsch--Mehr--Satz (KMS) potential~\cite{Karsch:1987pv}, while the real part of the potential is kept unchanged. We then solve the time-dependent Schr\"odinger equation for bottomonium evolution using the quantum trajectories method~\cite{Blaizot:2018oev, Brambilla:2021wkt, Omar:2021kra} in a realistic hydrodynamic background.

Using this framework, we compute nuclear modification factors $R_{AA}$ and several experimentally accessible double ratios for the bottomonium states $\Upsilon(1S)$, $\Upsilon(2S)$, and $\Upsilon(3S)$ in Pb--Pb collisions at $\sqrt{s_{NN}}=5.02$ TeV. Our results are compared with measurements from the ALICE, ATLAS, and CMS collaborations. We demonstrate that incorporating machine-learning--derived medium parameters leads to improved agreement with experimental data, particularly for excited bottomonium states that are most sensitive to medium-induced dissociation effects.
\section{QTraj Framework}
\label{sec02}
The in-medium real-time evolution of heavy quarkonium in this work is performed using QTraj~\cite{Omar:2021kra}, an open-source numerical solver designed to simulate heavy-quarkonium dynamics in the quark-gluon plasma using the quantum-trajectories approach. Before introducing the specific heavy-quark potentials employed in the present calculation, we briefly summarize the capabilities of the QTraj framework.

QTraj can solve Lindblad evolution equations derived from potential non-relativistic QCD (pNRQCD) combined with the open quantum systems (OQS) framework. In this approach, the heavy $Q\bar Q$ system contains both color-singlet and color-octet sectors, and its interaction with the thermal medium generates a non-Hermitian effective Hamiltonian together with collapse operators. QTraj can evolve the system either without quantum jumps, using only the effective Hamiltonian, or with stochastic jumps that generate transitions between singlet and octet color states. The framework can be applied to Lindblad equations obtained at leading order (LO) and next-to-leading order (NLO) in the binding-energy-to-temperature ratio, $E/T$~\cite{Brambilla:2017zei, Brambilla:2022ynh, Brambilla:2023hkw, Strickland:2024oat, Brambilla:2024tqg, Brambilla:2025sis, Thapa:2025jua}.

QTraj also supports complex-valued isotropic~\cite{Islam:2020gdv, Islam:2020bnp} and anisotropic~\cite{Islam:2025yxt} KSU potentials for heavy-quarkonium evolution in the QGP; a comprehensive list of supported potentials is provided in Appendix E of Ref.~\cite{Omar:2021kra}. The isotropic KSU potential describes an isotropic medium, while the anisotropic KSU potential incorporates the effects of momentum-space anisotropy. In contrast to the pNRQCD+OQS Lindblad evolution, for the KSU potentials the evolution is restricted to the color-singlet sector, and jumps or transitions between color-singlet and color-octet states are not considered.
\section{Isotropic KSU Potential Model}
\label{sec03}
In vacuum we take the heavy-quark potential to be given by a Cornell potential with a finite string breaking distance \cite{Islam:2020gdv, Islam:2020bnp, Islam:2025yxt}
\be
V_{\rm vac}(r) =
\begin{cases}  
	-\frac{a}{r} + \sigma r &\mbox{if } r \leq r_{\rm SB} \\
	-\frac{a}{r_{\rm SB} } + \sigma r_{\rm SB}   & \mbox{if } r > r_{\rm SB}
\end{cases} \, ,
\label{eq:vvac}
\ee
where \mbox{$a = 0.409$} is the effective coupling, \mbox{$\sigma = 0.21$~GeV$^2$} is the string tension, and \mbox{$r_{\rm SB}  =$ 1.25 fm} is the string breaking distance.  Using this set of vacuum parameters and assuming \mbox{$M_b = 4.7$ GeV} one obtains vacuum masses of \mbox{$\{9.46,10.0,9.88,10.36,10.25,10.13\}$ GeV} for $\Upsilon(1S)$, $\Upsilon(2S)$, $\chi_b(1P)$, $\Upsilon(3S)$, $\chi_b(2P)$, and $\Upsilon(1D)$ respectively. We have the internal energy based real part of the Karsch-Mehr-Satz (KMS) potential~\cite{Matsui:1986dk, Karsch:1987pv, Shuryak:2004tx, Strickland:2011aa}
\begin{equation}
\begin{aligned}
V_{\mathrm{KMS}}(r) ={}&
-\frac{a}{r}(1+m_D^R r)e^{-m_D^R r}
+\frac{2\sigma}{m_D^R}
    \left[1-e^{-m_D^R r}\right]
\\
&-\sigma r e^{-m_D^R r}\, ,
\end{aligned}
\label{rKMS}
\end{equation}
where the usual in-medium gluonic Debye mass is given by 
\begin{equation}
    m_D = \lambda \sqrt{\frac{4\pi}{3}N_C\left(1+\frac{N_F}{6}\right)\alpha_s T^2}
    \label{mD}
\end{equation}
In Eq.~(\ref{rKMS}), we take effective coupling as $a = \frac{g^2}{4\pi}C_F = \alpha_s C_F = \frac{4}{3}\alpha_s = 0.409$ and the Debye mass for the real part becomes
\begin{equation}
    m_D^R = \lambda \sqrt{\pi ~N_C\left(1+\frac{N_F}{6}\right)aT^2}~,
\end{equation}
where $N_C = 3,~ N_F = 2,~C_F = 4/3$, and $\lambda$ is an adjustable pre-factor called $m_D$-factor. Here, we will consider the case $\lambda = 1$, which is the known high-temperature limit of the gluonic Debye mass. To match smoothly onto the zero temperature limit, we use
\begin{equation}
V_R(r) =
\begin{cases}  V_{\rm KMS}(r)  &\mbox{if } V_{\rm KMS}(r)  \leq V_{\rm vac}(r_{\rm SB}) \\
	V_{\rm vac}(r_{\rm SB}) & \mbox{if } V_{\rm KMS}(r) > V_{\rm vac}(r_{\rm SB})
\end{cases} \, .
\label{eq:vmedre}
\end{equation}
In the limit $T\rightarrow0$, Eq.~\eqref{eq:vmedre} reduces to Eq.~\eqref{eq:vvac}. 

For the imaginary part of the potential, we take the result of the leading-order resummed perturbative QCD calculation of Laine et. al~\cite{Laine:2006ns}
\begin{equation}
V_I(r) = - C_F \alpha_s T \phi(m_D^I~ r) \, ,
\label{eq:vmedim}
\end{equation}
where
\begin{equation}
   \phi(\hat{r}) =  1 - 2 \hat{r}^2\int_0^\infty dz\frac{\sin(z)}{(z^2 + \hat{r}^2)^2}~,
\end{equation}
with $\hat{r} = r~m_D^I$. Here $m_D^I$ is given by the same as Eq.~(\ref{mD}) 
\begin{equation}
    m_D^I = \lambda \sqrt{\frac{4\pi}{3}N_C\left(1+\frac{N_F}{6}\right)\alpha_s T^2}
    \label{mDI}
\end{equation}
We evaluate the strong coupling $\alpha_s$ at the scale $\mu = 2 \pi T$ and use three-loop running \cite{pdg} with \mbox{$\Lambda_{\overline{MS}} = 344$ MeV}.  This value of $\Lambda_{\overline{MS}}$ is chosen in order to reproduce the lattice result for the running coupling $\alpha_s(5\text{ GeV}) = 0.2034$~\cite{McNeile:2010ji}.
The resulting total complex-valued potential is of the form 
\begin{equation}
V(r) = V_R(r) + i V_I(r) \, .
\label{eq:vform}
\end{equation}
The above phenomenological complex potential is known as an isotropic KSU potential model~\cite{Strickland:2011aa, Boyd:2019arx, Islam:2020gdv, Islam:2020bnp}. After solving the resulting evolution using the QTraj framework, we refer to the corresponding results as QTraj-Isotropic KSU throughout this work.
\section{The Machine Learning Framework}
\label{sec04}
In this work we employ a data-driven framework based on a deep-learning quasi-parton gas model (DLQPM), which allows one to extract temperature-dependent quasi-particle properties of the QGP directly from lattice QCD equation-of-state data~\cite{Li:2022ozl, Li:2025csc, HotQCD:2014kol}. In this approach, the strongly interacting plasma is described as a gas of quasi-particles whose effective masses depend on temperature and encode the non-perturbative interactions of the medium. These effective masses are determined by training deep neural networks to reproduce lattice QCD thermodynamic observables.

Once the temperature-dependent quasi-particle masses are obtained, they can be related to the running strong coupling constant and the Debye screening mass through perturbatively motivated relations~\cite{Jamal:2025gjy}. This provides a consistent way to incorporate non-perturbative lattice QCD information into the construction of the in-medium heavy-quark potential used in our quarkonium evolution calculations.

In the following subsections, we summarize the essential ingredients of this framework: the deep-learning quasi-parton gas model used to determine the quasi-particle masses, the extraction of the temperature-dependent strong coupling constant, and the determination of the Debye screening mass.
\subsection{Deep-Learning Quasi-Parton gas Model}
\begin{figure}[ht]
	\begin{center}
		\includegraphics[width=0.78\linewidth]{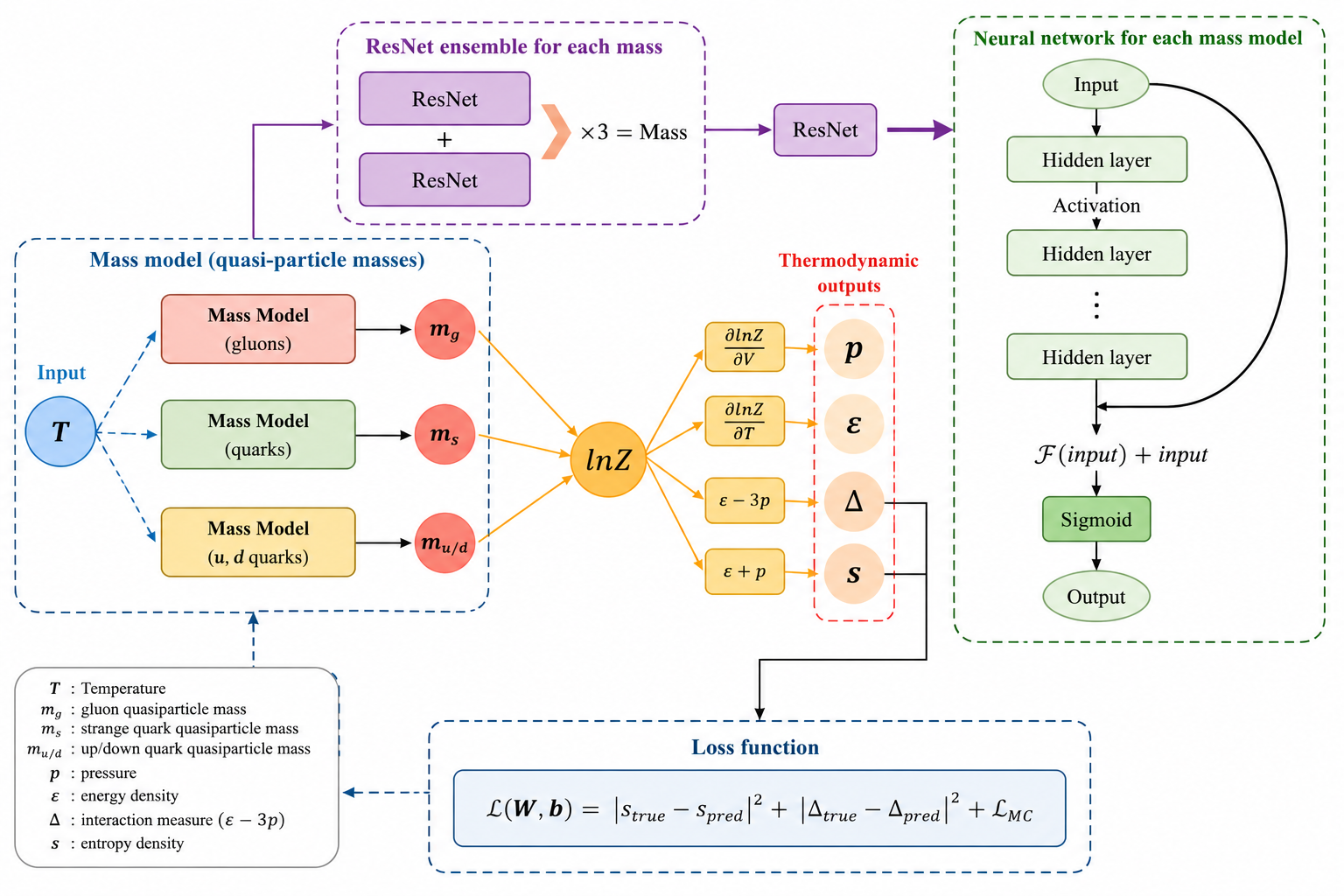}
	\end{center}
\caption{Schematic illustration of the deep-learning quasi-parton gas model (DLQPM) framework used to extract temperature-dependent quasiparticle masses and thermodynamic quantities from lattice QCD inputs.}
\label{DLQPM_framework}
\end{figure}

The starting point of the framework is the quasi-particle description of the QGP, in which the interacting system of quarks and gluons is approximated by an ideal gas of effective degrees of freedom with temperature-dependent masses~\cite{Li:2022ozl}. The framework of the neural network model is illustrated in Fig.~\ref{DLQPM_framework}. In this picture, the thermodynamic properties of the medium can be computed from the partition function of quasi-particles obeying Bose--Einstein and Fermi--Dirac statistics.
For a system consisting of gluons and quarks in thermal equilibrium, the total partition function can be written as~\cite{Kapusta_Gale_2006}
\begin{equation}
\ln Z(T) = \ln Z_g(T) + \sum_i \ln Z_{q_i}(T),
\end{equation}
where $Z_g$ represents the gluon contribution and $Z_{q_i}$ denotes the contribution of quarks with flavor $i$. The corresponding expressions are given by
\begin{equation}
\begin{split}
\ln Z_g(T) ={}&
-\frac{16V}{2\pi^2}
\int_0^\infty p^2\,dp \\
&\times
\ln\left[
1-\exp\left(
-\frac{\sqrt{p^2+m_g^2(T)}}{T}
\right)
\right].
\end{split}
\end{equation}
\begin{equation}
\begin{split}
\ln Z_{q_i}(T) ={}&
\frac{12V}{2\pi^2}
\int_0^\infty p^2\,dp \\
&\times
\ln\left[
1+\exp\left(
-\frac{\sqrt{p^2+m_{q_i}^2(T)}}{T}
\right)
\right].
\end{split}
\end{equation}
Here $V$ denotes the system volume and $p$ the quasi-particle momentum. The temperature-dependent quasi-particle masses $m_g(T)$ and $m_{q_i}(T)$ encode the medium interactions and modify the dispersion relation,
\begin{equation}
E(p,T) = \sqrt{p^2 + m_i^2(T)}.
\end{equation}
From the partition function, one obtains the thermodynamic quantities using the standard relations
\begin{equation}
P(T) = T \left( \frac{\partial \ln Z}{\partial V} \right)_T,
\qquad
\epsilon(T) = \frac{T^2}{V} \left( \frac{\partial \ln Z}{\partial T} \right)_V .
\end{equation}
The entropy density and trace anomaly are then given by
\begin{equation}
s(T) = \frac{\epsilon(T) + P(T)}{T},
\qquad
\Delta(T) = \epsilon(T) - 3P(T).
\end{equation}
In the DLQPM framework the unknown mass functions $m_g(T)$, $m_{u/d}(T)$, and $m_s(T)$ are represented by deep neural networks that take the temperature $T$ as input and output the corresponding quasi-particle masses. Each mass function is modeled using a residual neural network (ResNet) architecture with multiple hidden layers and nonlinear activation functions, allowing the network to learn smooth functional dependencies without imposing an explicit parametrization.

The neural-network parameters are determined by minimizing the deviation between the predicted thermodynamic quantities and lattice QCD results. The training objective typically involves the mean-squared error of the entropy density and trace anomaly together with additional physical constraints ensuring the correct high-temperature behavior of the quasi-particle masses. After training, the network provides temperature-dependent effective masses that reproduce lattice QCD thermodynamics over the relevant temperature range of the QGP.
\subsection{Determination of running coupling constant}
Once the temperature-dependent quasi-particle masses are obtained from the DLQPM framework, they can be used to determine the effective strong coupling constant of the medium. The overall procedure is illustrated in Fig.~\ref{DLQPM_framework}, where the temperature is provided as input to the neural network and the resulting quasi-particle masses are used to compute thermodynamic observables and extract the coupling constant.

The neural network simultaneously predicts three quasi-particle mass functions corresponding to gluons, light quarks, and strange quarks. These masses enter the partition function and allow the pressure, energy density, entropy density, and trace anomaly to be computed. The network parameters are optimized by minimizing the loss function~\cite{Li:2022ozl}
\begin{equation}
\mathcal{L} = (s_{\text{true}} - s_{\text{pred}})^2 + (\Delta_{\text{true}} - \Delta_{\text{pred}})^2 + \mathcal{L}_{\text{MC}},
\end{equation}
where $s_{\text{true}}$ and $\Delta_{\text{true}}$ denote lattice QCD results for entropy density and trace anomaly and $s_{\text{pred}}, \Delta_{\text{pred}}$ are the corresponding neural-network predictions. The term $\mathcal{L}_{\text{MC}}$ imposes additional constraints on the quasi-particle masses to ensure consistency with perturbative QCD behavior at high temperature.

The resulting quasi-particle masses can be related to the temperature-dependent gauge coupling through the leading-order thermal mass relations of QCD~\cite{Khvorostukhin:2010aj, Plumari:2011mk},
\begin{equation}
m_g^2(T) = \frac{1}{6} g^2(T) \left( N_c + \frac{1}{2} N_f \right) T^2,
\end{equation}
\begin{equation}
m_{u/d}^2(T) = \frac{N_c^2 - 1}{8N_c} g^2(T) T^2 .
\end{equation}
These relations allow the coupling constant to be extracted directly from the neural-network masses. Combining the two expressions yields
\begin{equation}
g^2(T) =
\frac{m_g^2(T) + m_{u/d}^2(T)}
{\left[\frac{1}{6}\left(N_c + \frac{1}{2}N_f\right) + \frac{N_c^2 - 1}{8N_c}\right] T^2}.
\end{equation}
The temperature-dependent strong coupling constant is then defined as
\begin{equation}
\alpha_s^{\text{ML}}(T) = \frac{g^2(T)}{4\pi}.
\label{mlalphas}
\end{equation}
This procedure provides a non-perturbative determination of $\alpha_s^{\text{ML}}(T)$ that is consistent with lattice QCD thermodynamics while preserving the perturbative structure of thermal mass relations.
\begin{figure}[ht]
	\begin{center}
		\includegraphics[width=0.75\linewidth]{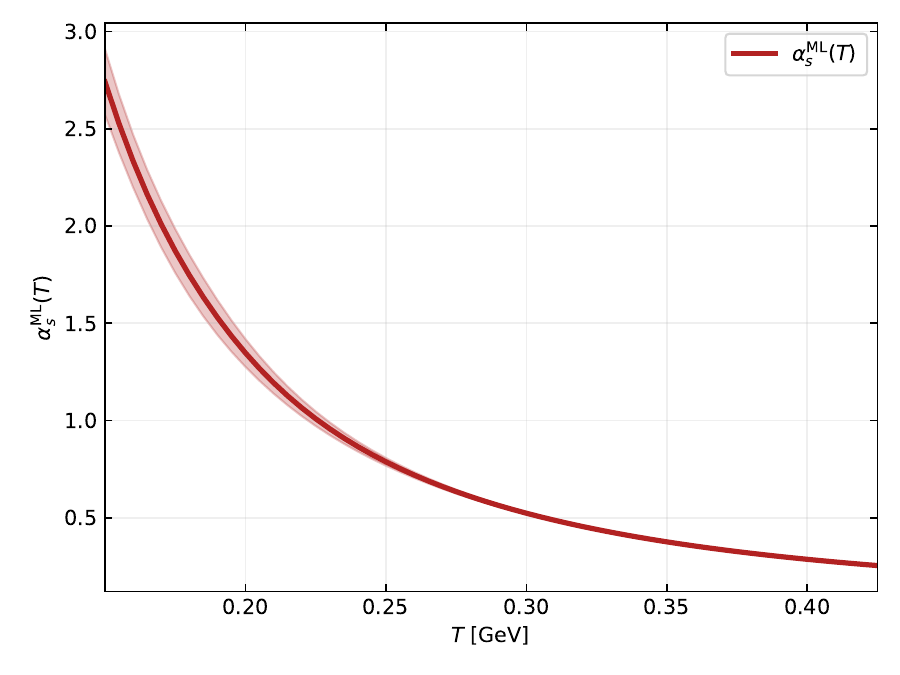}
	\end{center}
\caption{Temperature dependence of the strong coupling constant $\alpha_s^{\text{ML}}(T)$ extracted from the deep-learning quasi-parton gas model. The central curve represents the coupling obtained from the neural-network–determined quasi-particle masses using Eq.~(\ref{mlalphas}), while the shaded band indicates the uncertainty associated with the machine-learning extraction.}
\label{mlaplhasT}
\end{figure}
The temperature dependence of the running coupling constant obtained using this procedure is shown in Fig.~\ref{mlaplhasT}. The shaded uncertainty band reflects the variance obtained from multiple neural-network trainings used to estimate the uncertainty in the extracted quasi-particle masses. The coupling decreases monotonically with increasing temperature, consistent with the asymptotic freedom of QCD. We note that the extracted coupling becomes relatively large near the QCD crossover temperature. In this regime, the QGP is known to behave as a strongly coupled medium, and the quantity $\alpha_s^{\text{ML}}(T)$ obtained here should therefore be interpreted as an effective medium coupling derived from lattice-constrained quasi-particle masses rather than a perturbative coupling evaluated at a fixed renormalization scale.
\subsection{Debye mass in HTL framework}
In thermal QCD, the Debye screening mass arises from the static limit of the longitudinal gluon self-energy in a hot medium. Within the Hard Thermal Loop (HTL) approximation, the Debye mass is defined as the static limit of the temporal component of the gluon polarization tensor \cite{Braaten:1990it,Braaten:1991gm,Blaizot:2001nr},
\begin{equation}
m_D^2 = -\Pi_{00}(k_0=0,\mathbf{k}\rightarrow 0),
\label{eq:Debye_definition}
\end{equation}
where $\Pi_{00}$ denotes the temporal component of the gluon polarization tensor. 

Evaluating the one-loop gluon self-energy in a thermal medium leads to the following kinetic-theory representation of the Debye mass \cite{Bellac:2011kqa, Kapusta:2006pm, Jamal:2018mog},
\begin{equation}
\begin{aligned}
m_D^2(T) = -4\pi\alpha_s(T)\Bigg[
&2N_c \int \frac{d^3q}{(2\pi)^3}
\frac{\partial f_g(q,T)}{\partial q} \\
&+ 2N_f \int \frac{d^3q}{(2\pi)^3}
\frac{\partial f_q(q,T)}{\partial q}
\Bigg].
\end{aligned}
\label{HTLmD}
\end{equation}
In the quasiparticle description employed in this work, the thermal distributions contain the temperature-dependent effective masses extracted from the DLQPM. Accordingly, the gluon and quark equilibrium distribution functions are written as
\begin{equation}
f_g(q, T)=\frac{1}{e^{E_g(q,T)/T}-1},
\qquad
f_q(q, T)=\frac{1}{e^{E_q(q,T)/T}+1},
\end{equation}
where,
\begin{equation}
E_g(q, T)= \sqrt{q^2+m^2_g(T)},
\qquad
E_q(q, T)=\sqrt{q^2+m^2_q(T)},
\end{equation}
In the massless limit, $m_g, m_q \rightarrow 0$, these distributions reduce to the standard equilibrium Bose–Einstein and Fermi–Dirac distributions with $E(q) =q $. In this limit, the momentum integrals in Eq. (\ref{HTLmD}) can be evaluated analytically, yielding the familiar leading-order HTL expression for the Debye screening mass \cite{Kapusta:2006pm, Laine:2016hma}
\begin{equation}
m_D^{\mathrm{LO}}(T)=T\sqrt{4\pi\alpha_s(T)\left(\frac{N_c}{3}+\frac{N_f}{6}\right)}.
\label{lo}
\end{equation}
%
\subsection{Determination of ML Debye mass}
Once the temperature-dependent coupling $\alpha_s^{\mathrm{ML}}(T)$ is obtained, it is substituted into 
Eq.~(\ref{HTLmD}) to compute the machine learning Debye screening mass,
\begin{equation}
\begin{aligned}
m_D^{\mathrm{ML}}(T) = \Bigg[-4\pi\alpha_s^{\mathrm{ML}}(T)\Bigg(
&2N_c \int \frac{d^3q}{(2\pi)^3}
\frac{\partial f_g(q,T)}{\partial q} \\
&+ 2N_f \int \frac{d^3q}{(2\pi)^3}
\frac{\partial f_q(q,T)}{\partial q}
\Bigg)\Bigg]^{1/2}.
\end{aligned}
\label{MLmD}
\end{equation}
This formulation preserves the kinetic-theory structure of the Debye screening mass while incorporating the lattice-constrained, machine-learning–extracted quasiparticle properties through both the effective coupling $\alpha_s^{\text{ML}}(T)$ and the temperature-dependent quasiparticle masses entering the distribution functions.
\begin{figure}[ht]
	\begin{center}
		\includegraphics[width=0.75\linewidth]{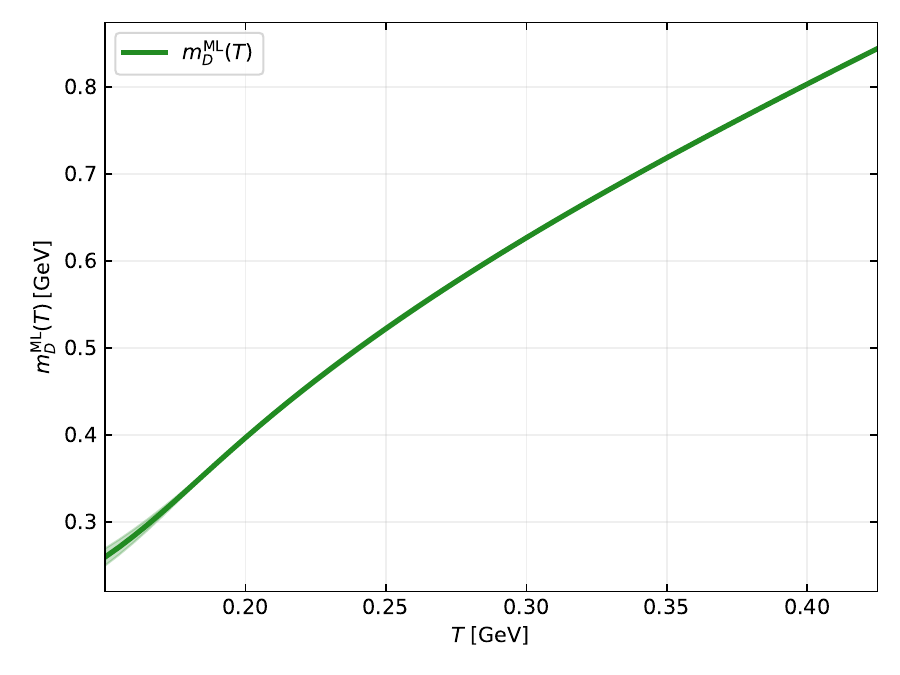}
	\end{center}
\caption{Temperature dependence of the Debye screening mass $m_D^{\text{ML}}(T)$ obtained from the machine-learning--derived running coupling using Eq.~(\ref{MLmD}).}
\label{mlmd}
\end{figure}
The temperature dependence of the Debye screening mass $m_D^{\text{ML}}(T)$ obtained using Eq.~(\ref{MLmD}) is shown in Fig.~\ref{mlmd}. As expected from thermal QCD, the screening mass increases with temperature due to the rising density of quarks and gluons in the medium. Physically, this behavior reflects the enhanced screening of color interactions in the quark–gluon plasma, which reduces the range of the heavy-quark potential at higher temperatures. The resulting Debye mass therefore provides the characteristic screening scale governing the modification of heavy-quark bound states in the hot medium.
\subsection{Effective ML Debye Mass}
In the HTL framework, the Debye mass satisfies $m_D(T) \propto \sqrt{\alpha_s(T)}\,T$, indicating that variations in the effective strong coupling should be accompanied by a corresponding rescaling of the screening mass. Within our machine-learning framework, both the temperature-dependent coupling $\alpha_s^{\mathrm{ML}}(T)$ and the Debye mass $m_D^{\mathrm{ML}}(T)$ are obtained from the quasi-parton model trained on lattice QCD thermodynamics. However, in the imaginary part of the KSU potential the coupling is evaluated at the thermal scale $2\pi T$~\cite{Strickland:2011aa, Boyd:2019arx, Islam:2020gdv, Islam:2020bnp}, while the machine-learning coupling is determined at the scale $T$. To preserve the HTL scaling relation while maintaining consistency with the scale used in the potential, we therefore define an effective machine-learning Debye mass
\begin{equation}
    m_{D}^{\text{ML,eff}}(T) = m_D^{\text{ML}}(T) \sqrt{\frac{\alpha_s(2\pi T)}{\alpha_s^{\text{ML}}(T)}}
    \label{MLeff}
\end{equation}
This procedure ensures that the screening scale entering the Landau-damping contribution retains the correct parametric dependence on the strong coupling while incorporating the non-perturbative information contained in the machine-learning determination of medium properties.
\begin{figure}[ht]
	\begin{center}
		\includegraphics[width=0.75\linewidth]{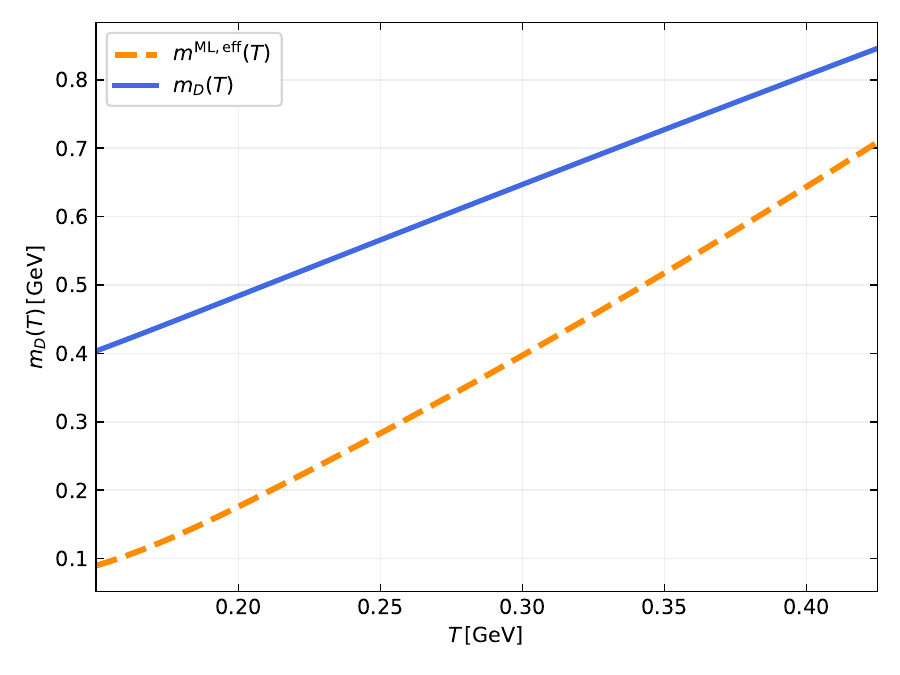}
	\end{center}
\caption{Comparison between the theoretical Debye screening mass $m_D(T)$ obtained from Eq.~(\ref{lo}) (solid blue curve) and the effective machine learning Debye mass $m_D^{\mathrm{ML,eff}}(T)$ obtained from Eq.~(\ref{MLeff}) (dashed orange curve).}
\label{mleffmd}
\end{figure}
Figure~\ref{mleffmd} compares the Debye screening mass obtained from the theoretical HTL-inspired expression (\ref{lo}) with the effective Debye mass (\ref{MLeff}) extracted within the machine-learning quasi-particle framework. Both quantities exhibit a monotonic increase with temperature, consistent with the enhanced screening of color interactions in the quark--gluon plasma as the density of thermal partons grows. However, the ML-derived effective Debye mass remains smaller than the perturbative estimate at lower temperatures, indicating the presence of non-perturbative medium effects captured by the data-driven quasi-particle masses. At higher temperatures the two curves approach each other, reflecting the gradual transition toward perturbative QCD behavior.
\subsection{ML induced potential model}
We replace the in-medium gluonic Debye mass, $m_D$, with the machine learning based effective Debye mass, $m_D^{\mathrm{ML,eff}}(T)$, in the imaginary part of the potential, i.e., Eq. (\ref{eq:vmedim})
\begin{equation}
V_I^{\text{ML}}(r) = - C_F \alpha_s T \phi(m_{D}^{\text{ML,eff}}~ r) \, .
\label{MLVI}
\end{equation}
In the present implementation, this modification is applied only to the imaginary part of the potential, while the real part is kept identical to that of the isotropic KSU model. The reason is that the imaginary component directly encodes in-medium scattering and decoherence processes that are sensitive to the screening scale, whereas the real part primarily controls the binding structure and is already well constrained within the KSU framework. This separation allows us to isolate the impact of machine-learning-inferred medium properties on the in-medium decay width of quarkonium states while maintaining compatibility with the established description of the binding potential.

Finally, we take the machine learning based total complex potential of the form
\begin{equation}
V^{\text{ML}}(r) = V_R(r) + i V^{\text{ML}}_I(r) \, .
\label{mlt}
\end{equation}
The above complex-valued potential (\ref{mlt}) is referred to as the ML-Central China Normal University (CCNU) potential model. After solving the resulting evolution using the QTraj framework, we refer to the corresponding predictions as QTraj-ML CCNU throughout this work.
\section{Numerical method for solving the Schr\"odinger equation}
\label{sec05}
Since the complex heavy-quark potential considered in this work is spherically symmetric, the time-dependent Schrödinger equation can be decomposed into time-dependent radial equation and time-independent angular momentum channels. Defining the reduced radial wave function $u_\ell(r,t)=rR_\ell(r,t)$, the Hamiltonian for each partial wave is
\begin{equation}
\hat{H}_\ell = \frac{\hat{p}^2}{2m} + V_{\mathrm{eff},\ell}(r,t),
\qquad
V_{\mathrm{eff},\ell}(r,t) = V(r,t) + \frac{\ell(\ell+1)}{2mr^2},
\label{eq:radial_hamiltonian}
\end{equation}
where $\hat{p}=-i\partial/\partial r$. The corresponding real-time evolution is given by
\begin{equation}
u_\ell(r,t+\Delta t) = \exp\left(-i\hat{H}_\ell\Delta t\right) u_\ell(r,t).
\label{eq:uUpdate}
\end{equation}
We solve Eq.~\eqref{eq:uUpdate} using the split-step pseudospectral method developed and described in detail in Ref.~\cite{Boyd:2019arx}; see also Refs.~\cite{Islam:2020bnp, Omar:2021kra, Dong:2022mbo, Islam:2025yxt} for detailed implementations of the method. In this approach, the time-evolution operator is implemented using a symmetric operator splitting between the effective-potential and kinetic-energy contributions. The kinetic step is evaluated in momentum space using discrete sine transforms (DSTs), which automatically enforce the radial boundary condition $u_\ell(0,t)=0$. The DSTs are efficiently evaluated using Fast Fourier Transform (FFT) routines, and the numerical evolution is massively parallelized on GPUs using the CUDA CUFFT library~\cite{cuda}.

For the present calculation, the procedure is applied independently to the $\ell=0$ and $\ell=1$ channels relevant for the bottomonium states considered below. Further details of the numerical implementation, convergence properties, and split-step evolution algorithm can be found in Ref.~\cite{Boyd:2019arx}.
\section{Computation of \texorpdfstring{$R_{AA}$}{RAA}~including feed-down}
\label{sec06}
We solve the time-dependent Schr\"odinger equation using the complex potentials of Eq.~\eqref{eq:vform} and  Eq.~\eqref{mlt} numerically on a discrete radial lattice. The real-time evolution is performed with a split-step pseudospectral scheme based on discrete sine transforms (DST) \cite{Fornberg:1978,TAHA1984203,Boyd:2019arx}. The system is discretized using $N=4096$ points with $r_{\rm max}=19.7$ fm, resulting in a lattice spacing of approximately $0.0048$ fm. The initial condition is taken to be a narrow Gaussian\cite{Islam:2020gdv}
\be
u_\ell(r,\tau=0) \propto r^{\ell+1} \exp(-r^2/\Delta^2) \, ,
\ee
with $\Delta=0.04$ fm. This localized initial wave packet models the short-distance production of the heavy-quark pair in a hard scattering. Following the in-medium evolution, the survival probability for each bottomonium state is determined by projecting the evolved wave function onto the corresponding vacuum eigenstate.

We solve the 3+1D Schrödinger equation for quarkonium states on a realistic $3+1$D anisotropic-hydrodynamic (aHydroQP) background calibrated to $\sqrt{s_{NN}}=5.02$ TeV heavy-ion collision data~\cite{Alqahtani:2020paa,Alqahtani:2017mhy}. The hydrodynamic evolution is initialized using smooth optical Glauber conditions, with a central temperature $T_0=630$ MeV at the initial proper time $\tau_0=0.25$ fm/$c$ and a specific shear viscosity corresponding to $4\pi\eta/s=2$~\cite{Alqahtani:2020paa}. Prior to the medium formation time $\tau_{\mathrm{med}}$, the quarkonium wave packets propagate according to the vacuum potential. For $\tau\geq\tau_{\mathrm{med}}$, the appropriate in-medium complex potential is used along each trajectory as long as the local temperature remains above $T_{\mathrm{QGP}}=155$ MeV. Once the local temperature falls below this value, the evolution is continued using the vacuum potential.

To account for the spatial and temporal variation of the medium temperature, we perform the time-dependent Schrödinger evolution for a sample of one million bottomonium trajectories. The initial transverse production coordinates $(x,y)$ are sampled according to the nuclear binary-collision overlap profile, $N_{AA}^{\mathrm{bin}}(x,y)$. Bottomonium states are initialized at midrapidity $(y=0)$, with transverse momenta drawn from the distribution $p_T/(p_T^2+\langle M\rangle^2)^2$ and azimuthal angles randomly selected within $\phi\in[0,2\pi]$. The local QGP temperature experienced by each sampled state along its straight-line trajectory is obtained from the aHydroQP background. After propagating each state along its trajectory, the survival probabilities are used to determine the effective production cross sections by multiplying them by the average number of binary collisions in the selected centrality bin and the primordial production cross section of each state.  

Feed-down contributions from higher-lying bottomonium states are included through a feed-down matrix $F$, whose elements are constructed using the branching fractions reported by the Particle Data Group~\cite{ParticleDataGroup:2020ssz}:
\begin{equation}
	F_{ij} = \left\{ \begin{matrix}
		\text{branching fraction $j$ to $i$}, & \text{for } i < j, \\
		1, & \text{for } i = j, \\
		0, & \text{for } i > j,
		\end{matrix} \right.\label{eq:fdmnew}
\end{equation}
which accounts for decays of excited states into lower states (see also eq.~(6.4) of ref.~\cite{Brambilla:2020qwo}). A detailed construction of the individual matrix elements \eqref{eq:fdmnew} is given in \cite{Islam:2020bnp}.

For p-p collisions, the primordial and experimentally observed post-feed-down production cross sections are related through $\vec{\sigma}_\text{exp} = F \vec{\sigma}_\text{primordial}$. Our considered nine bottomonium states are $\vec{\sigma} = \{ \Upsilon(1S),\,$ $\Upsilon(2S),\,$ $\chi_{b0}(1P),\,$ $\chi_{b1}(1P),\,$ $\chi_{b2}(1P),\,$ $\Upsilon(3S),\,$ $\chi_{b0}(2P),\,$ $\chi_{b1}(2P),\,$ $\chi_{b2}(2P)\}$. Using the post-feed-down production cross sections $\vec{\sigma}_{\text{exp}}=\{57.6$, 19, 3.72, 13.69, 16.1, 6.8, 3.27, 12.0, $14.15\}$ nb, the corresponding primordial cross sections are obtained by inverting the feed-down relation, $\vec{\sigma}_\text{primordial} = F^{-1} \vec{\sigma}_\text{exp}$.
In heavy-ion collisions, the post-QGP effective production cross section of each state is determined using $\Sigma_{\mathrm{QGP},i} =
S_i \left\langle N_{\mathrm{bin}}(b) \right\rangle \sigma_{\mathrm{primordial},i}$ where $S_i$ is the survival probability of state $i$ and $\left\langle N_{\mathrm{bin}}(b) \right\rangle$ is the average number of binary collisions. The final effective production cross sections, including feed-down, are then obtained from $\vec{\Sigma}_{\mathrm{final}}=F\vec{\Sigma}_{\mathrm{QGP}}$. Finally, the nuclear suppression factor, $R_{AA}^i$ , for state $i$ is evaluated as $R_{AA}^{i}=\Sigma_{\mathrm{final},i}/ \left(\left\langle N_{\mathrm{bin}}\right\rangle
\sigma_{\mathrm{exp},i}\right)$, where $\left\langle N_{\mathrm{bin}}\right\rangle$ corresponds to the selected centrality class and $\sigma_{\mathrm{exp},i}$ is the associated post-feed-down $pp$ production cross section.
\section{Results}
\label{sec07}
\begin{figure*}[ht!]
	\begin{center}
		\includegraphics[width=0.485\linewidth]{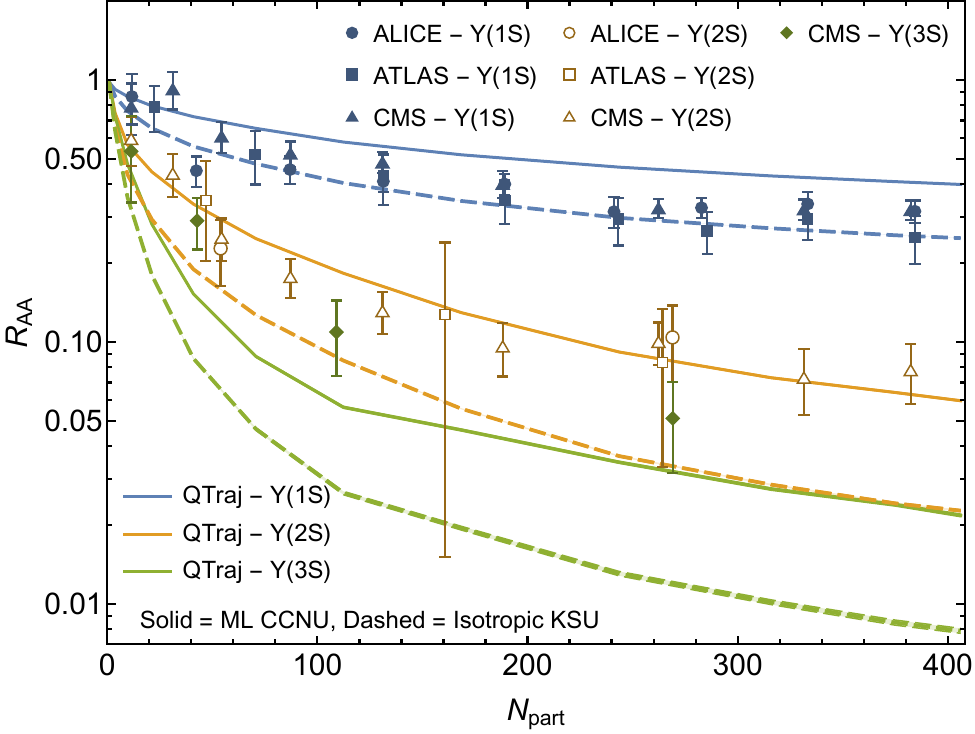}\hspace{2mm}
		\includegraphics[width=0.485\linewidth]{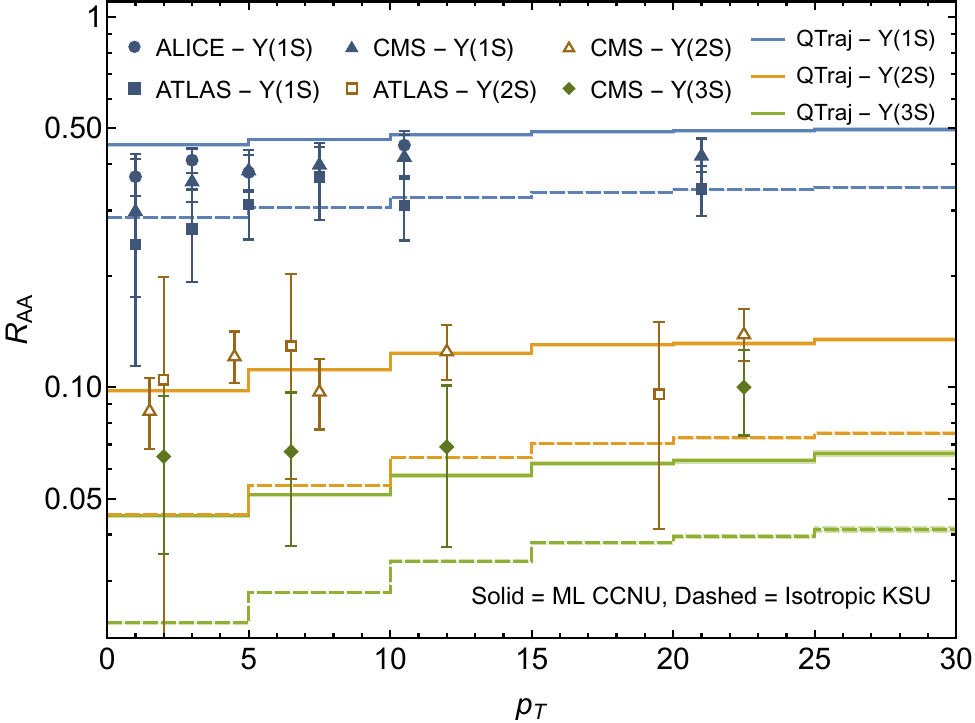}
	\end{center}
	\caption{Nuclear suppression factor, $R_{AA}$, of bottomonium s-wave states as a function of $N_\text{part}$ (left panel) and $p_T$ (right panel). The solid and dashed lines indicate results obtained with ML CCNU and isotropic KSU potential models, respectively. The experimental measurements shown are from the ALICE~\cite{ALICE:2020wwx}, ATLAS~\cite{ATLAS5TeV}, and CMS~\cite{Sirunyan:2018nsz,CMS-PAS-HIN-21-007} collaborations. Experimental error bars shown were obtained by adding statistical and systematic uncertainties in quadrature.}
	\label{raa}
\end{figure*}

Figure~\ref{raa} shows the nuclear modification factor $R_{AA}$ for the bottomonium $\Upsilon(1S)$, $\Upsilon(2S)$, and $\Upsilon(3S)$ states in Pb--Pb collisions at $\sqrt{s_{NN}} = 5.02$ TeV as a function of the number of participating nucleons $N_{\text{part}}$ (left panel) and the transverse momentum $p_T$ (right panel). The solid curves correspond to results obtained using the ML-CCNU potential within the QTraj framework, while the dashed curves show predictions based on the conventional isotropic KSU potential. Experimental measurements from the ALICE~\cite{ALICE:2020wwx}, ATLAS~\cite{ATLAS5TeV}, and CMS~\cite{Sirunyan:2018nsz,CMS-PAS-HIN-21-007} collaborations are included for comparison. 

A clear sequential suppression pattern is observed, with the ground state $\Upsilon(1S)$ being the least suppressed and the excited states $\Upsilon(2S)$ and $\Upsilon(3S)$ exhibiting progressively stronger suppression as the medium density increases. While both models reproduce this qualitative hierarchy, the ML-CCNU model provides a noticeably improved quantitative description of the experimental data, particularly for the excited states. In the isotropic KSU model the suppression of $\Upsilon(2S)$ and $\Upsilon(3S)$ is generally overestimated, especially for central collisions, resulting in $R_{AA}$ values that fall significantly below the measurements. In contrast, the ML-CCNU results predict larger survival probabilities for these states and follow the experimental trends more closely across the full centrality range. A similar improvement is also visible in the $p_T$ dependence shown in the right panel, where the ML-CCNU curves remain systematically closer to the experimental data. This improvement can be traced to the machine-learning-induced modification of the Debye screening mass entering the imaginary part of the heavy-quark potential, which directly controls the in-medium dissociation rate of quarkonium states. Since excited states are more weakly bound and therefore more sensitive to medium-induced damping, the ML-based Debye mass leads to a more realistic estimate of their thermal decay widths, resulting in better agreement with the observed suppression pattern. These results demonstrate that machine-learning-informed medium parameters provide a promising and systematic way to incorporate lattice-constrained QCD information into phenomenological models of quarkonium suppression in the quark-gluon plasma.

\begin{figure*}[ht!]
	\begin{center}
		\includegraphics[width=0.485\linewidth]{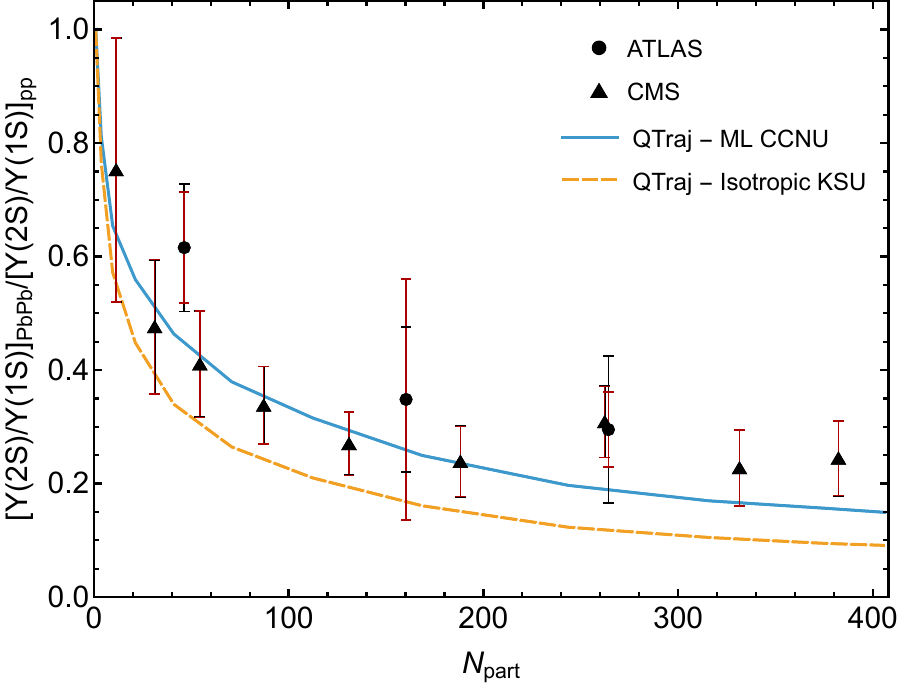}\hspace{2mm}
		\includegraphics[width=0.485\linewidth]{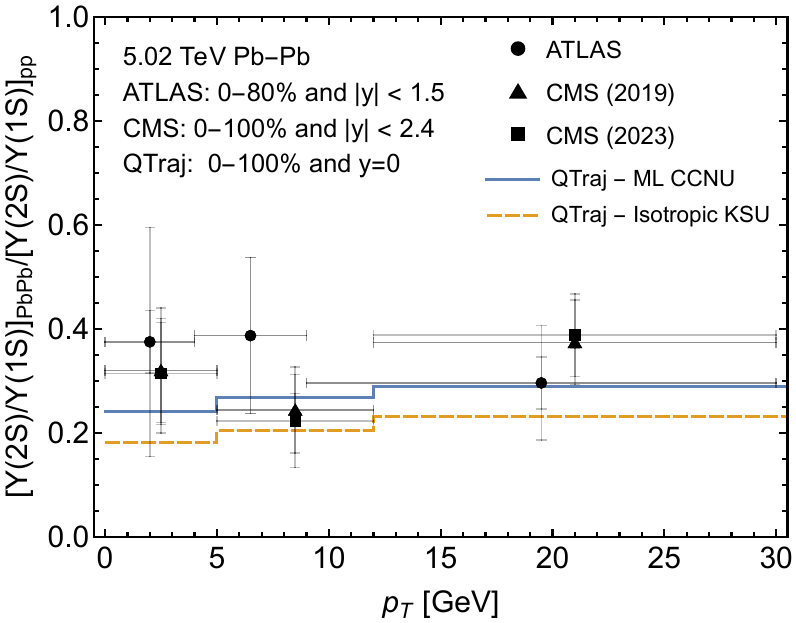}
	\end{center}
	\caption{Double ratio $[\Upsilon(2s)/\Upsilon(1s)]_\text{PbPb}/[\Upsilon(2s)/\Upsilon(1s)]_\text{pp}$ as a function of $N_\text{part}$ (left panel) and $p_T$ (right panel). The solid and dashed lines indicate results obtained with ML CCNU and isotropic KSU potential models, respectively. The experimental measurements shown are from the ATLAS~\cite{ATLAS5TeV} and CMS~\cite{Sirunyan:2018nsz, CMS-PAS-HIN-21-007} collaborations. }
	\label{2s1s}
\end{figure*}
The double ratio $[\Upsilon(2S)/\Upsilon(1S)]_{\text{PbPb}}/[\Upsilon(2S)/\Upsilon(1S)]_{\text{pp}}$ is shown in Fig.~\ref{2s1s} as a function of the number of participating nucleons $N_{\text{part}}$ (left panel) and the transverse momentum $p_T$ (right panel). Experimental measurements from the ATLAS~\cite{ATLAS5TeV} and CMS~\cite{Sirunyan:2018nsz, CMS-PAS-HIN-21-007} collaborations are shown for comparison. For the ATLAS measurement of the $2S$ to $1S$ double ratio, the black and red error bars represent the statistical and systematic uncertainties, respectively. The data points labeled CMS (2023) were obtained by reconstructing the $2S$ to $1S$ double ratio from the published $p_T$-dependent measurements of $R_{AA}[1S]$ and $R_{AA}[2S]$. In contrast, for the ATLAS and CMS (2019) results, the collaborations reported the $2S$ to $1S$ double ratio directly. The double ratio provides a particularly sensitive observable for probing the relative suppression of excited bottomonium states with respect to the ground state, largely reducing uncertainties associated with the initial production cross sections. In the left panel, the double ratio decreases with increasing $N_{\text{part}}$, reflecting the stronger suppression of the more weakly bound $\Upsilon(2S)$ state in increasingly central collisions. While both models reproduce this qualitative behavior, the ML-CCNU predictions remain systematically closer to the experimental measurements across the full centrality range. In contrast, the isotropic KSU model predicts a substantially stronger suppression, leading to double-ratio values that fall significantly below the data. The same trend is observed in the $p_T$ dependence shown in the right panel, where the ML-CCNU results track the measured values more closely than the isotropic KSU predictions. This improvement originates from the machine-learning--induced modification of the Debye screening mass entering the imaginary part of the heavy-quark potential, which reduces the effective in-medium decay width of excited states and leads to a more realistic description of their relative survival probability.

\begin{figure*}[ht!]
	\begin{center}
		\includegraphics[width=0.550\linewidth]{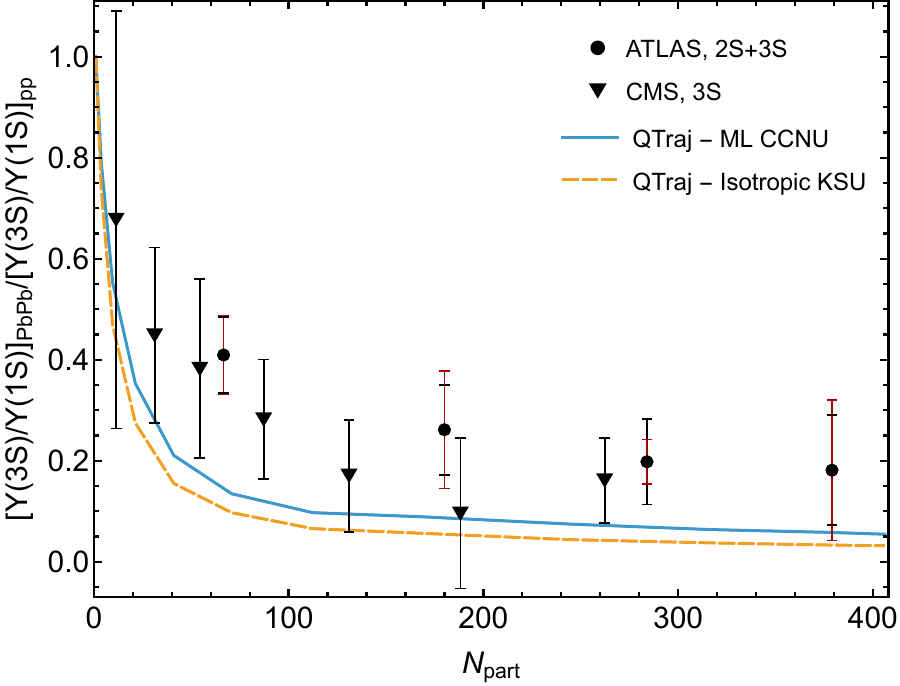}
	\end{center}
	\caption{Double ratio $[\Upsilon(3S)/\Upsilon(1S)]_\text{PbPb}/[\Upsilon(3S)/\Upsilon(1S)]_\text{pp}$ as a function of $N_\text{part}$. Line styles and experimental data sources are the same as Fig.~\ref{2s1s}.}
	\label{3s1s}
\end{figure*}
\begin{figure*}[ht!]
	\begin{center}
		\includegraphics[width=0.485\linewidth]{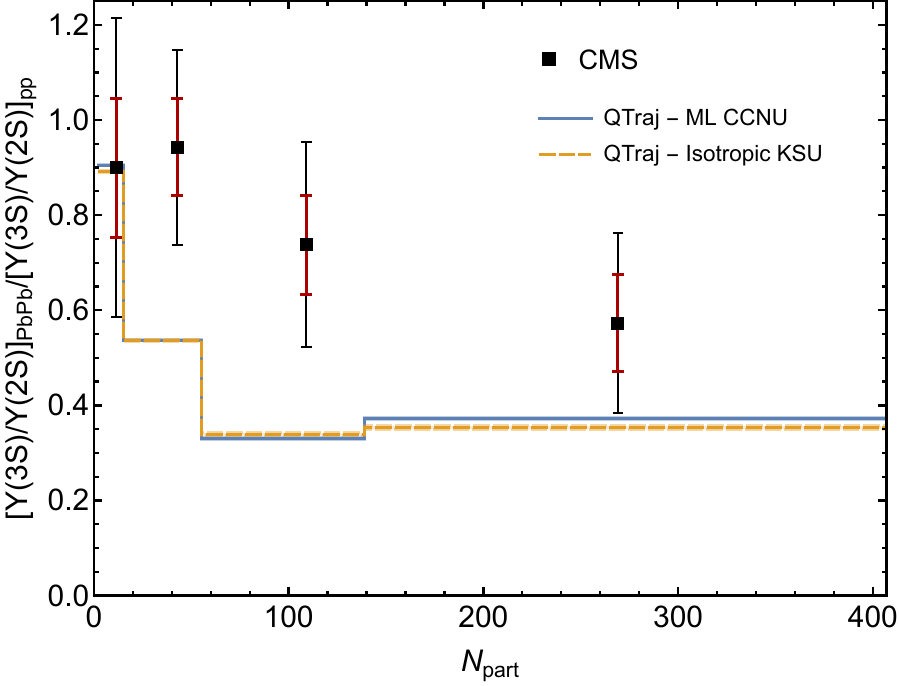}\hspace{2mm}
		\includegraphics[width=0.485\linewidth]{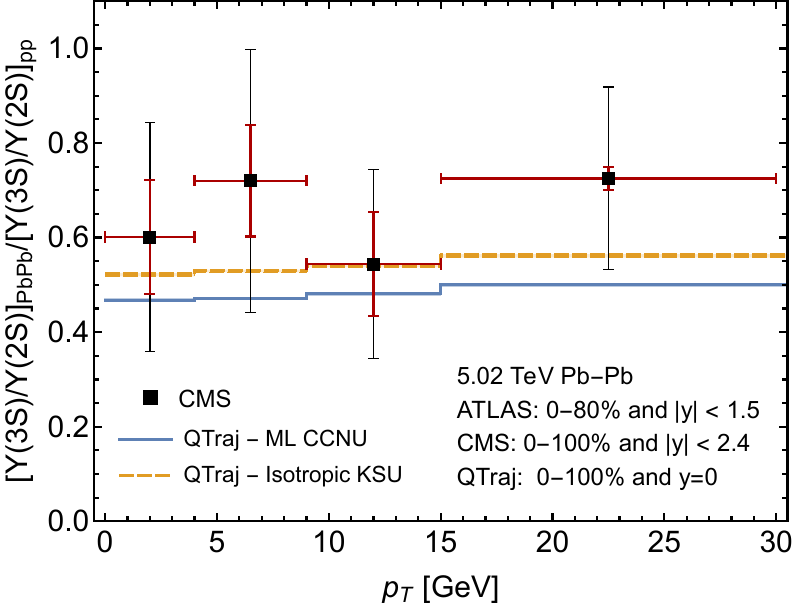}
	\end{center}
	\caption{Double ratio $[\Upsilon(3S)/\Upsilon(2S)]_\text{PbPb}/[\Upsilon(3S)/\Upsilon(2S)]_\text{pp}$ as a function of $N_\text{part}$ (left panel) and $p_T$ (right panel). Line styles are the same as Fig.~\ref{2s1s}.  The centrality classes used were 0-30\%, 30-50\%, 50-70\%, and 70-90\%.   Experimental data are from Ref.~\cite{CMS-PAS-HIN-21-007}.}
	\label{3s2s}
\end{figure*}

The double ratio $[\Upsilon(3S)/\Upsilon(1S)]_{\text{PbPb}}/[\Upsilon(3S)/\Upsilon(1S)]_{\text{pp}}$ is shown in Fig.~\ref{3s1s} as a function of the number of participating nucleons $N_{\text{part}}$. For the CMS $3S$ to $1S$ double ratio, we reconstruct this observable using the $3S$ to $2S$ double ratio reported by CMS (shown in Fig.~\ref{3s2s}) together with the inferred $2S$ to $1S$ double ratio presented in Fig.~\ref{2s1s}. In the case of ATLAS, we instead use the experimentally reported combined $(2S + 3S)$ to $1S$ double ratio~\cite{Sirunyan:2018nsz}. Since the $\Upsilon(3S)$ state is the most weakly bound among the bottomonium s-wave states considered in this work, this observable provides a particularly sensitive probe of the in-medium dissociation dynamics. As expected, the double ratio decreases rapidly with increasing $N_{\text{part}}$, reflecting the strong suppression of the $\Upsilon(3S)$ state in more central collisions. While both models capture the qualitative trend, the ML-CCNU predictions remain consistently closer to the experimental data over the full centrality range. In contrast, the isotropic KSU model tends to predict a stronger suppression, leading to smaller double-ratio values that underestimate the measurements. The improved agreement obtained with the ML-CCNU model is again attributable to the machine-learning--derived Debye mass entering the imaginary part of the heavy-quark potential, which modifies the in-medium decay width of quarkonium states. Because the $\Upsilon(3S)$ state is particularly sensitive to medium effects due to its small binding energy, the ML-based medium parameters lead to a more realistic description of its survival probability.

The double ratio $[\Upsilon(3S)/\Upsilon(2S)]_{\text{PbPb}}/[\Upsilon(3S)/\Upsilon(2S)]_{\text{pp}}$ is shown in Fig.~\ref{3s2s} as a function of the number of participating nucleons $N_{\text{part}}$ (left panel) and the transverse momentum $p_T$ (right panel). This observable probes the relative suppression between the two excited bottomonium states and therefore provides a sensitive test of the sequential suppression pattern in the quark--gluon plasma. In the left panel, the double ratio decreases with increasing $N_{\text{part}}$, reflecting the stronger suppression of the more weakly bound $\Upsilon(3S)$ state in more central collisions. The ML-CCNU prediction follows the overall centrality trend of the data slightly better than the isotropic KSU model, which tends to predict somewhat smaller values. In contrast, for the $p_T$ dependence shown in the right panel, the isotropic KSU model appears to lie marginally closer to the experimental measurements across most $p_T$ bins, while the ML-CCNU prediction remains somewhat lower. Overall, the two models yield similar predictions for this observable, indicating that the relative suppression between the two excited states is less sensitive to the detailed form of the in-medium potential than the ratios involving the ground state, where the ML-CCNU model showed clearer improvements.
\section{Conclusions and outlook}
\label{sec08}
In this work, we have investigated bottomonium suppression in heavy-ion collisions using a machine-learning--informed description of the in-medium heavy-quark potential. Specifically, we introduced a machine-learning--derived Debye screening mass, obtained from a deep-learning quasi-parton gas model constrained by lattice QCD thermodynamics, into the imaginary part of the complex Karsch--Mehr--Satz (KMS) potential. The resulting complex potential was then used to evolve bottomonium wave packets in a realistic quark--gluon plasma background using the quantum trajectories (QTraj) framework. Within this setup, we computed the nuclear modification factors $R_{AA}$ and several double ratios for the bottomonium states $\Upsilon(1S)$, $\Upsilon(2S)$, and $\Upsilon(3S)$ in Pb--Pb collisions at $\sqrt{s_{NN}} = 5.02$ TeV and compared our predictions with measurements from the ALICE, ATLAS, and CMS collaborations.

Our results demonstrate that incorporating machine-learning--derived medium parameters leads to a noticeably improved description of experimental data, particularly for excited bottomonium states. For the nuclear modification factors, the ML-CCNU model predicts larger survival probabilities for the $\Upsilon(2S)$ and $\Upsilon(3S)$ states compared to the conventional isotropic KSU potential, resulting in significantly better agreement with the measured suppression pattern as a function of both centrality and transverse momentum. This improvement originates from the modified Debye screening mass entering the imaginary part of the heavy-quark potential, which directly controls the in-medium decay width of quarkonium states. Since excited states are more weakly bound and therefore more sensitive to medium-induced damping, the ML-based description leads to a more realistic estimate of their dissociation dynamics in the quark--gluon plasma.

We further examined several experimentally measured double ratios that probe the relative suppression among bottomonium states. In particular, the observables\\
$[\Upsilon(2S)/\Upsilon(1S)]_{\text{PbPb}}/[\Upsilon(2S)/\Upsilon(1S)]_{\text{pp}}$ and $[\Upsilon(3S)/\Upsilon(1S)]_{\text{PbPb}}/[\Upsilon(3S)/\Upsilon(1S)]_{\text{pp}}$ are well described by the ML-CCNU model and show clear improvement relative to the isotropic KSU predictions. For the ratio involving only excited states, $[\Upsilon(3S)/\Upsilon(2S)]$, both models produce similar results, indicating that this observable is less sensitive to the detailed form of the in-medium potential than ratios involving the ground state.

Taken together, our results demonstrate that machine-learning--informed medium properties provide a powerful and systematic way to incorporate lattice-constrained QCD information into phenomenological models of quarkonium suppression. Embedding such data-driven inputs into dynamical frameworks such as QTraj opens a new pathway toward improving the quantitative description of heavy quarkonium evolution in the quark--gluon plasma.

Several directions for future work remain open. First, it would be interesting to extend the present framework by incorporating machine-learning--derived medium properties into both the real and imaginary parts of the heavy-quark potential. Second, applying this approach to other quarkonium systems, such as charmonium, could provide additional tests of the role of data-driven medium parameters in quarkonium dissociation. Third, the present study focused on isotropic medium effects; incorporating anisotropic plasma dynamics within the ML-based potential framework may further improve the description of observables such as quarkonium elliptic flow. More broadly, this work demonstrates the potential of combining modern machine-learning techniques with real-time quantum evolution methods to develop next-generation theoretical frameworks for describing strongly interacting matter in heavy-ion collisions, which can be further tested and constrained by future high-precision measurements at the LHC and upcoming facilities such as the Electron--Ion Collider.
\vspace{3mm}
\section*{Data Availability}
The data and code supporting this study are available from the corresponding repositories~\cite{qtraj-download, DLQPM:FPLi}. 

Additional data are available from the authors upon reasonable request.
\section*{Acknowledgments}
The authors wish to thank Mohammad Yousuf Jamal for useful discussions. A.I., S.C., and G.Y.Q. were supported in part by the Natural Science Foundation of China (NSFC) under grant No. 12225503. F.P. L. acknowledges support from
the National Natural Science Foundation of China (NSFC) under Grant Nos. 12325507, 12547102, and 12147101, from the National Key Research and Development Program of China under Grant No. 2022YFA1604900, and in part from the China Postdoctoral Science Foundation under Grant No. 2025M783370. L.G.P. was supported in part by the Natural Science Foundation of China (NSFC) under grant No. 12435009. This work was also supported by the Outstanding Leading Talent Team Program of Central China Normal University (XJ2026000302). Numerical computations were performed on the high-performance computing facilities of the Nuclear Science Computing Center at Central China Normal University (NSC$^3$).
\bibliography{main}
\end{document}